\documentclass[pre,twocolumn,aps,eqsecnum]{revtex4}
\usepackage{epsfig}

\begin{document}

\title{Thermodynamic theory of dislocation-mediated plasticity}

\author{J.S. Langer}
\affiliation{Dept. of Physics, University of California, Santa
Barbara, CA  93106-9530}

\author{Eran Bouchbinder}
\affiliation{Racah Institute of Physics, Hebrew University of Jerusalem, Jerusalem 91904, Israel}

\author{Turab Lookman}
\affiliation{Theoretical Division, Los Alamos National Laboratory, Los Alamos, NM 87544}

\date{\today}
\begin{abstract}
We reformulate the theory of polycrystalline plasticity, in externally driven, nonequilibrium situations, by writing equations of motion for the flow of energy and entropy associated with dislocations.  Within this general framework, and using a minimal model of thermally assisted depinning with essentially only one adjustable parameter, we find that our theory fits the strain-hardening data for Cu over a wide range of temperatures and six decades of strain rate. We predict the transition between stage II and stage III hardening, including the observation that this transition occurs at smaller strains for higher temperatures.  We also explain why strain-rate hardening is very weak up to large rates; and, with just one additional number, we accurately predict the crossover to power-law rate hardening in the strong-shock regime. Our analysis differs in several important respects from conventional dislocation-mediated continuum  theories.  We provide some historical background and discuss our rationale for these differences.

\end{abstract}
\maketitle

\section{Introduction}
\label{intro}

\subsection{Historical Background}

The time seems ripe for a critical reexamination of the theory of dislocation-mediated plasticity.  The basic elements of dislocation theory were described decades ago in the classic books by Cottrell \cite{COTTRELL-53}, Friedel \cite{FRIEDEL-67}, and Hirth and Lothe \cite{HIRTH-LOTHE-68}. Since then, major progress has been made via numerical simulations and experimental observations; but several fundamental issues remain unresolved. Most notably, a first-principles theory of strain hardening has yet to be developed.  

The question of whether such a theory is feasible remains a matter of serious debate.  The prevailing opinion among experts in the field is that it is not.  For example, Cottrell \cite{COTTRELL-02} has argued recently that strain hardening [rather than turbulence], is ``the most difficult remaining problem in classical physics.''  He goes on to explain that ``neither of the two main strategies of theoretical many-body physics -- the statistical mechanical approach; and the reduction of the many-body problem to that of the behaviour of a single element of the assembly -- is available to work [strain] hardening. The first fails because the behaviour of the whole system is governed by that of weakest links, not the average, and is thermodynamically irreversible. The second fails because dislocations are flexible lines, interlinked and entangled, so that the entire system behaves more like a single object of extreme structural complexity and deformability, that Nabarro and I once compared to a bird's nest, rather than as a set of separate small and simpler elementary bodies.'' 

Similarly, on page 235 of their definitive review of strain hardening, Kocks and Mecking \cite{KOCKS-MECKING-03} tell us pessimistically that ``an ab initio theory of strain hardening, with a quantitative prediction of the numerical constants, is unlikely to ever be derived even for a specific case, and impossible with any generality.'' They advocate what, from a physicist's point of view, is a purely phenomenological approach, based on extensive observations and a search for trends, but with no hope of uncovering fundamental principles that might lead to predictive theories.  

Devincre, Hoc and Kubin  \cite{KUBIN-08} are equally pessimistic when they say that ``The present dislocation-based models for strain hardening still have difficulties integrating elementary dislocation properties into a continuum description of bulk crystals or polycrystals.  As a consequence, current approaches cannot avoid making use of extensive parameter fitting.''  These authors then show that they can reproduce observed deformation curves by defining a dislocation mean free path and incorporating it, with input from numerical dislocation dynamics, into a multiscale analysis.  Unfortunately, their work is based on a well known equation of motion for the dislocation density that that we find problematic on basic physical grounds.  (See Sec.\ref{SH1}.)

If a first-principles theory of dislocation-mediated plasticity is intrinsically impossible, then this central part of solid mechanics is an intrinsically unfortunate research area -- much less fortunate, for example, than semiconductor electronics or polymer science, where basic understanding has been translated into quantitative theories which, in turn, have guided enormously successful engineering applications.  All of the authors cited above are keenly aware of this problem, but insist that nothing can be done about it.  In the present paper, we challenge their pessimism by proposing a different approach based on thermodynamic principles.  We examine Cottrell's rationales for failure -- especially the thermodynamically irreversible nature of dislocation dynamics -- from this broader point of view, and look to see where specific dislocation mechanisms play roles within a bigger picture. 

\subsection{Thermodynamics and Dislocations}

Cottrell's remark about thermodynamic irreversibility reflects a general consensus that thermodynamics, especially the second law, is irrelevant in dislocation theory.  It is well known that the energies of dislocations are so large that ordinary thermal fluctuations are completely ineffective in creating or annihilating them.  It is also well known that the entropy of dislocations is extremely small in comparison with the total entropy of the material that contains them.  (For example, see \cite{COTTRELL-53}, p.39.)  This happens because the dislocations involve only a very small fraction of all the atoms in the material and, therefore, account for only a very small fraction of the degrees of freedom of the system as a whole.  

The central assertion of this paper is that, although the dislocation entropy is small, it is an essential ingredient of a theory of dislocation-mediated deformation.  In writing equations of motion for a system that contains many millions of irregularly moving dislocations per square centimeter, it is necessary to require that these equations always take the system toward states of higher, and never lower, probability. Then, in circumstances where ordinary thermal fluctuations are irrelevant, the dislocation entropy by itself must be a non-decreasing function of time; and there must exist some dislocation-specific form of the second law of thermodynamics to describe this situation.  

The natural way in which to develop a second-law analysis for dislocations is to express the theory in terms of an ``effective temperature,'' denoted here by $\chi$ to distinguish it from the temperature $T$ that characterizes the ordinary thermal fluctuations.  Given a macroscopic sample of a material, we can compute the energy, say $U_C$, of any configuration of dislocations, and we can count the number of such configurations in any energy interval.  Thus we can compute a configurational entropy $S_C(U_C)$.  Following Gibbs, we define the effective temperature, in energy units, to be
\begin{equation}
\label{chidef}
\chi = {\partial U_C\over\partial S_C}.
\end{equation}
Because the dislocation energies are large, $\chi$ is an extremely large temperature, vastly greater than $T$. Therefore, although $S_C$ may be small, the product $\chi\,S_C$ is of the order of the dislocation energy $U_C$; and the state that minimizes the free energy $U_C-\chi\,S_C$ is the most probable state of the system.  An externally driven system, such as a deforming crystal, must be either in that most probable state or moving toward it. 

We emphasize that we are talking {\it only} about externally driven systems.  In the usual Gibbsian statistical thermodynamics, one considers equilibrium states in which thermal fluctuations cause systems to explore statistically significant fractions of their state spaces. If the dislocations are to explore their state spaces in the absence of such fluctuations, they must be driven to do so by external forces.  Accordingly, our thermodynamic theory of dislocations pertains only to situations in which there are external forces that persistently drive the system from one state to another, and produce complex, chaotic motions.  It is only in such cases that it makes sense to enforce the second law of effective-temperature thermodynamics. The resulting theory is well suited for experiments at very high shear rates, and also should work for most constant strain rate experiments under laboratory conditions.  But it is not relevant to quasistatic experiments such as indentation or beam bending; nor does it have anything to say (so far as we know) about strain-gradient theories, geometrically necessary dislocations, and the like. 

By using the word ``entropy'' in the same paragraph as the term ``driven system,'' we raise yet another controversial issue.   The question is whether the term ``nonequilibrium thermodynamics'' makes any sense at all; or, more specifically, whether it is conceptually possible to derive the classic Kelvin-Planck or Clausius statements of the second law directly from  Gibbsian principles of statistical physics. A large body of literature, starting with the papers of Coleman, Noll, and Gurtin in the 1960's \cite{COLEMAN-NOLL-63,COLEMAN-GURTIN-67}, neatly avoids this issue by adopting the Clausius-Duhem inequality as an axiomatic basis for a systematic formulation of nonequilibrium thermodynamics.  

Two of the present authors (EB and JSL in \cite{BLI-09,BLII-09,BLIII-09}) have found  that the axiomatic approach is not sufficient for nonequilibrium problems in solid mechanics, where one must deal with internal variables such as densities of flow defects or dislocations, and where the internal degrees of freedom may fall out of equilibrium with the thermal reservoir.  By carefully defining the entropy of an externally driven system containing internal variables, we are able to constrain the equations of motion for those variables to be consistent with the fundamental, statistical statement of the second law.  Our procedure underlies all of the following analysis. 

Finally, we emphasize that our thermodynamic formulation is not in any sense a replacement for the dislocation theories described  in \cite{COTTRELL-53,FRIEDEL-67,HIRTH-LOTHE-68} and in many texts and research reports published since those classic works appeared.  As will be seen in what follows, our analysis leads to the definition of a relatively small number of parameters describing, for example, dislocation production rates, or the rates at which external work is converted to configurational entropy in various circumstances.  To compute such quantities from first principles, we eventually will need to invoke all of the specific mechanisms that we know to be relevant -- the dynamics of different kinds of dislocations, their interactions with each other and with other internal structures, and the mechanisms by which they are created and destroyed. We expect that, in future research, we will be better able to understand these mechanisms once they are seen in a broader thermodynamic context.  

\subsection{Scheme of This Paper}

In order to make progress beyond thermodynamics, we need a physics-based model that relates the plastic strain rate and dislocation motion to the driving stress.  In Sec.\ref{Model-dynamics}, we propose a minimal model of dislocation pinning and depinning that begins to address Cottrell's concerns about the irregular, irreversible, collective behavior of dislocations.  Our model describes thermally activated deformation during stage-III hardening and dynamic recovery, and it also provides a starting point for describing stage-II hardening.  (See \cite{KOCKS-MECKING-03} for a definition of the several ``stages'' of strain hardening.) However, it does not describe the deterministically chaotic motion of dislocations seen at very low temperatures and extremely small strain rates.  That behavior is discussed in detail by Zaiser \cite{ZAISER-06} who, in the concluding part of his review, makes it clear that our range of interest does not (yet) overlap with his.  

Section \ref{Teff} is devoted to the details of our thermodynamic analysis.  The ideas discussed there have emerged from recent developments in the shear-transformation-zone (STZ) theory of amorphous plasticity \cite{BLIII-09,FL-98,JSL-STZ-PRE-08}, where the key ingredient is an effective  temperature that characterizes the internal state of disorder.  The effective temperature in the STZ theory is exactly the same as the temperature $\chi$ defined in Eq.(\ref{chidef}).  Both STZs and dislocations are configurational defects; that is, they are irregularities in the underlying atomic structures of solid materials. In comparison to the ephemeral STZs, however, dislocations are long-lived, spatially extended objects.  Because dislocation energies are so much greater than thermal energies, and because the dislocations do not participate in glasslike jamming transitions and thus need no internal degrees of freedom of their own, the effective temperature analysis is conceptually simpler for dislocation-mediated plasticity in crystalline solids than it is for deformations of glassy materials. We have taken advantage of this simplicity by writing Sec.\ref{Teff} in a language that we intend to be self explanatory. 

A major impetus for the present investigation has been the paper by Preston, Tonks, and Wallace (PTW) \cite{PTW-03} on plasticity at extreme loading conditions.  These authors construct a phenomenological formula for the stress as a function of strain, strain rate, and temperature, for a number of elemental metallic solids.  Their formula fits direct experimental measurements and data deduced indirectly from shock tests, for strain rates ranging from moderate to explosively large, and and for temperatures ranging from cryogenic up to substantial fractions of melting points.  It describes plastic flow as a thermally activated process for strain rates $\dot\epsilon$ up to at least $10^4\,\,{\rm sec}^{-1}$.  In the explosively fast regime, for strain rates above about $10^8\,\,{\rm sec}^{-1}$, the PTW curves of stress $\sigma$ {\it versus} strain rate cross over to a power law, $\sigma \sim \dot\epsilon^{\beta}$, with $\beta \cong 0.25$. The scope of the PTW analysis, plus the apparently universal nature of the phenomena that they describe, led us to suspect that there might be a more fundamental way to understand these behaviors.  We use the PTW data for Cu in Secs.\ref{SH2} and \ref{VHSR} for comparisons between our theory and experiment.

Sections \ref{SH1}, \ref{scaling}, and \ref{SH2} contain our analysis of strain hardening.  We start in Sec.\ref{SH1} by explaining why the storage-recovery equation of Kocks and coworkers \cite{KOCKS-66,MECKING-KOCKS-81,FOLLANSBEE-KOCKS-88,KOCKS-MECKING-03} is unsuitable for our purposes, and then go on to derive an alternative equation based on the second law of thermodynamics and energy conservation. The resulting theory allows us only one fully adjustable parameter, specifically, the temperature (but not strain-rate) dependent fraction of the external work that is converted to configurational heat, as opposed to being stored in the form of dislocations or other internal structures. Numerical results and comparisons with experiment are presented in Sec.\ref{SH2}.  We show there that our theory fits the strain-hardening data for Cu over a wide range of temperatures and six decades of strain rate.  It predicts the transition between stage II and stage III hardening, and  explains why this transition occurs earlier at higher temperatures.  It also explains why strain-rate hardening is very weak up to large rates. Finally, in Sec.\ref{VHSR}, we show how a simple extension of these thermodynamic ideas accounts for the observed crossover to strong, power-law, rate hardening at extremely high strain rates in the strong-shock regime. \cite{PTW-03}  

We conclude in Sec.\ref{summary} with a brief summary of our results and remarks about future directions for investigation.  We emphasize that equations of the form proposed here are well suited for the study of position dependent instabilities such as shear banding, and we look ahead to comparing those equations to the ones used by Ananthakrishna \cite{ANANTHAKRISHNA-07} in his studies of dynamic dislocation  patterns.

\section{Minimal Model of Dislocation Dynamics} 
\label{Model-dynamics}

The minimal model to be used here is a polycrystalline material subject to a simple shear stress of magnitude $\sigma$. Our key assumption is that dislocation motion is controlled entirely by a thermally activated depinning mechanism.  Thus we start with a model that we expect to be most relevant to stage-III hardening and dynamic recovery.  We will see in Sec.\ref{SH1} how this model can be used in a theory that includes athermal, stage II behavior. 

We assume that, after coarse-grained averaging, our system is spatially homogeneous and orientationally symmetric; orientational symmetry is broken only by the direction of the applied stress.  We further assume that we can describe the population of dislocations by a single, averaged, areal density $\rho$.  For more detailed applications, $\rho$ will need to be replaced by a set of such densities, each describing different kinds of dislocations with different orientations, and we will need separate equations of motion for each of these populations. Even in that case, however, we can assume that these populations retain their identities and do not, like STZs, transform dynamically from one internal state to another, thus requiring additional internal degrees of freedom.  As a result, generalizing to a many-population version of the present theory should not raise fundamentally new issues. 

Suppose that the population of dislocations moves with average velocity $v$ in response to $\sigma$.  The plastic shear rate is given by the Orowan relation
\begin{equation}
\label{Orowan}
\dot\epsilon^{pl} = \rho\,b\,v;~~~~\rho = \ell^{-2},
\end{equation}
where $b$ is the magnitude of the Burgers vector and $\ell$ is the average distance between dislocations.  In the spirit of leaving no assumption unquestioned, note that Eq.(\ref{Orowan}) is a purely geometric relation.  Each dislocation produces a shear $b/L$ as it moves across a system of linear size $L$.  There are $\rho\,L^2$ dislocations doing this at any time; and the rate at which these events are occurring is $v/L$. The product of these factors is the right-hand side of Eq.(\ref{Orowan}). Strictly speaking, both sides of this equation are tensors; but we suppress the tensor notation in interests of simplicity. 

The next step is to compute the velocity $v$.  To do this, assume that each dislocation moves through an array of pinning sites, which may be a forest of cross dislocations.  Alternatively, the pinning sites might be grain boundaries, point defects, stacking faults, or some combination of all of these.  Because a dislocation is a long-lived entity, it becomes pinned and unpinned many times during its motion across the system.  Here we make a minor departure from conventional dislocation theory.  Rather than distinguishing between mobile and pinned (or ``stored'') dislocations, we prefer to assume that each dislocation retains its identity throughout these processes. 

If a dislocation spends a characteristic time $\tau_P$ at a pinning site, and then (as in \cite{KUBIN-08}) moves almost instantaneously across some mean free path $\ell^*$ before being trapped again, its average speed is
\begin{equation}
v_P(\sigma) = {\ell^*\over \tau_P(\sigma)}. 
\end{equation}
To estimate $v_P(\sigma)$, assume a thermally activated  mechanism in which the dislocation is trapped in a potential well of depth $U_P(0) \equiv k_B T_P$ in the absence of an external stress, and can escape from this trap via a thermal fluctuation.  The activation temperature $T_P$ may be large, in fact, greater than the melting temperature; but it remains much smaller than the dislocation formation energy. When a stress $\sigma$ is applied, the barrier opposing escape from the trap is lowered in the direction of $\sigma$ and raised in the opposite direction.  Since we need to evaluate $v_P$ for arbitrarily large $\sigma$, we cannot make the usual linear approximation in $\sigma$ for this barrier-lowering effect, but instead must let the barrier decrease smoothly toward zero.  The simplest way to do this is to write
\begin{equation}
\label{UP}
U_P(\sigma) = k_B T_P\,e^{-\sigma/\sigma_T},
\end{equation}
where $\sigma_T$ is a characteristic depinning stress.  

It seems clear that $\sigma_T$ must be the Taylor stress:
\begin{equation}
\label{Taylor}
\sigma_T = \mu_T\,{b\over \ell} = \mu_T\,b\,\sqrt{\rho},
\end{equation}
where $\mu_T$ is proportional to the shear modulus $\mu$. The point here is that, except in the case of an isolated dislocation in an exceptionally pure crystal, the array of elastically interacting dislocations impedes the internal motions.  The right-hand side of Eq.(\ref{Taylor}) is an estimate of the shear stress needed to depin one dislocation segment by moving it a distance, say $b'$, from its pinning site, where $\mu_T\,b/\ell = \mu\,b'/\ell$. We guess that the ratio $b'/b$ is of the order of $0.1$, so that $\mu_T \sim 0.1\,\mu$.  On dimensional grounds, it is hard to see how any other combination of parameters could play a role in Eq.(\ref{UP}).  

The velocity $v_P(\sigma)$ is 
\begin{equation}
\label{vP}
v_P(\sigma)= {\ell^*\over \tau_0}\,\Bigl[f_P(\sigma)-f_P(-\sigma)\Bigr],
\end{equation}
where $\tau_0^{-1}$ is a microscopic attempt frequency, of the order of $10^{12}\,{\rm sec}^{-1}$, and 
\begin{equation}
\label{fP}
f_P(\sigma) = \exp\left(- {T_P\over T}\,e^{-\sigma/\sigma_T}\right).
\end{equation}
Antisymmetry is required in Eq.(\ref{vP}) both to preserve reflection symmetry, and to satisfy the second-law requirement that the energy dissipation rate, $\sigma\,v_P(\sigma)$, is non-negative.  In almost all the situations to be considered here, the second term on the right-hand side of Eq.(\ref{vP}) is completely negligible compared to the first for positive $\sigma$. 

Note that this minimal model already addresses some of the concerns expressed by Cottrell.  Our use of the Taylor stress in Eq.(\ref{UP}) recognizes that the depinning mechanism is a collective effect, depending on the average spacing of the extended objects in the Cottrell-Nabarro ``bird's nest.''  Moreover, our assumption that the average velocity $v(\sigma)$ is accurately determined by only the depinning time $\tau_P$ is consistent with Cottrell's ``weakest link'' picture.  The speed at which a dislocation segment jumps from one pinning site to another is increasingly unimportant the greater that speed becomes; the slowest process -- depinning in this case -- is always the dominant one.  It is this property of the system that gives us latitude in choosing which of various competing mechanisms determines the barrier height $T_P$.  

To make the depinning argument more precise, suppose that the speed of an unpinned dislocation, say $v_D(\sigma)$, is linearly proportional to the Peach-Koehler force:
\begin{equation}
\label{vD}
v_D(\sigma) = {b\,\sigma\over \eta\,\tau_0},
\end{equation}
where $\eta$ is a drag coefficient with the dimensions of stress.  The motion of unpinned dislocations can make an appreciable difference in the total strain rate only if the drag time $\tau_D = \ell^*/v_D$ is at least as long as the pinning time $\tau_P$, that is, if $\ell^*\,\eta\,\tau_0/\,b\,\sigma \ge \tau_0/\,f_P(\sigma)$. In principle, this inequality might be satisfied for stresses that are so large that the activation rate saturates, $f_P(\sigma)\to 1$, but that are still less than $\eta$.  As will be seen, however, $f_P(\sigma)$ remains small because, at high strain rates, the density of dislocations $\rho$, and thus the Taylor stress $\sigma_T$, become large. Therefore, it is not necessary to invoke a high-stress, viscosity dominated regime, even at the highest strain rates reported in PTW.  On the other hand, the depinning argument, and the minimal model itself, do break down in Zaiser's \cite{ZAISER-06} limit of very low temperature and vanishingly small strain rates.  At $T=0$, there must be a finite stress that can overcome the pinning barrier, and the zero-temperature limit of $\tau_P$ cannot be infinite as is predicted by Eq.(\ref{fP}).  As Zaiser has shown, dislocation dynamics in that regime is very different than that described here.  

For all of the analysis that follows, we assume that we are not in the extreme low-temperature regime, so that $v = v_P(\sigma)$.  We also assume that the mean free path $\ell^*$ is proportional to $\ell$, and simply let $\ell^*=\ell$.  (A numerical factor here can be absorbed into the definitions of other parameters.)  Then we write 
\begin{equation}
\label{qdef}
q(\sigma,\rho) \equiv \dot\epsilon^{pl}\,\tau_0 = b\,\sqrt{\rho}\,\Bigl[f_P(\sigma)-f_P(-\sigma)\Bigr].
\end{equation}
The dimensionless plastic strain rate $q(\sigma,\rho)$ compares rates of mechanical deformation with intrinsic microscopic rates such as atomic vibration frequencies.  Ordinarily, laboratory stress-strain curves are measured at $q \ll 1$.  However, the highest rates in PTW approach $q \sim 1$, as do strain rates in fracture or in sheared granular materials. We discuss the high strain rate regime in Sec.\ref{VHSR}. 

Before going any further, it is useful to solve Eq.(\ref{qdef}) for the ratio $\sigma/\sigma_T$.  Except in the elastic region, at vanishingly small strains, the total strain rate is entirely plastic to a very good approximation, and $\sigma/\sigma_T$ is easily large enough that the reverse-stress term on the right-hand side of Eq.(\ref{qdef}) is negligible for positive stress.  Dropping that term, we find 
\begin{equation}
\label{sigma-sigmaT}
{\sigma\over \sigma_T} \approx \ln\left({T_P\over T}\right) - \ln\left[{1\over 2}\,\ln\left({b^2\,\rho\over \,q^2}\right) \right]\equiv \nu(T,\,\rho,\,q)
\end{equation}
Note that $q$ appears in Eq.(\ref{sigma-sigmaT}) only as the argument of a double logarithm; thus rate hardening for $q \ll 1$ is extremely slow, as is observed experimentally.  Most importantly, $\nu(T,\,\rho,\,q)$, is a very slowly varying function of all its arguments, indicating that $\sigma$ is always a slowly varying multiple of the Taylor stress within the range of validity of this approximation. 

\section{Effective-Temperature Thermodynamics}
\label{Teff}

The constitutive relation in Eq.(\ref{qdef}) must be supplemented by a theory that, ultimately, will tell us how to evaluate the dislocation density $\rho$ that appears there both explicitly and as the argument of the Taylor stress $\sigma_T$.  The thermodynamic ideas that we use to develop this theory are best -- in fact, necessarily -- expressed in terms of the effective temperature $\chi$ defined in Eq.(\ref{chidef}).  These ideas are discussed in detail in \cite{BLII-09}; the results presented in \cite{JSL-MANNING-TEFF-07} are especially relevant to what follows.  For simplicity, we focus here only on the aspects of the effective temperature theory that are relevant to the dislocation problem.

We start by assuming that the internal degrees of freedom of a solidlike material can be separated into two, weakly interacting, subsystems.  The first, configurational, subsystem is defined by the mechanically stable positions of the constituent atoms, i.e. the ``inherent structures'' of Stillinger and Weber.\cite{STILLINGER-WEBER-82,STILLINGER-88} The second, kinetic-vibrational, subsystem is defined by the momenta and the displacements of the atoms at small distances away from their stable positions.  

In general, the kinetic-vibrational degrees of freedom are fast variables; they relax to equilibrium on atomic time scales.  As a result, they equilibrate rapidly with a heat bath, and are always at the bath temperature $T$.  On the other hand, the atomic rearrangements that take the configurational subsystem from one inherent structure to another are relatively slow and/or infrequent. For example, the generation of dislocations by Frank-Read sources is an extremely slow process in comparison with atomic vibration frequencies. The crucial point is that, when the dynamic coupling between the two subsystems is weak -- as is the coupling between dislocations and thermal fluctuations -- then the configurational degrees of freedom can be out of equilibrium with the kinetic-vibrational ones.  In that case, under circumstances specified more carefully in \cite{BLII-09}, the configurational subsystem has an effective temperature of its own, denoted here by $\chi$.  

For crystalline solids, the inherent structures are specified by the populations and positions of dislocations, grain boundaries, and other defects.  The energy $U_C$ introduced in the paragraph preceding Eq.(\ref{chidef}) is most accurately defined as the energy of an inherent structure.  Let ${\cal N}(U_C)\,dU_C$ be the number of such structures with energies in the interval $dU_C$, and define the dimensionless entropy to be $S_C = \ln\,({\cal N}/{\cal N}_0)$, where ${\cal N}_0$ is an irrelevant normalization constant.  Then $\chi = \partial U_C/\partial S_C$ has the dimensions of energy.  

Because $\chi$ is a thermodynamic temperature in the usual statistical sense, the most probable density of dislocations at given $\chi$ is proportional to a Boltzmann factor in the limit where this density is small, that is, in the limit in which the probability of any given lattice site being occupied by a dislocation is much less than unity.  Therefore,
\begin{equation}
\label{rho-chi}
\rho^{ss}(\chi) =  {1\over a^2}\,e^{-\,e_D/\chi},
\end{equation}
where $a$ is a length scale of the order of atomic spacings, and $e_D$ is a characteristic formation energy for dislocations -- the energy per unit length of a dislocation multiplied by some mesoscopic length such as the circumference of a dislocation loop or a typical grain size.  (See \cite{BLI-09,BLII-09} for a discussion of how this familiar formula emerges in a nonequilibrium context.) The superscript ``$ss$'' means that $\rho = \rho^{ss}(\chi)$ only when the system has undergone enough deformation that the dislocation density has reached a steady-state quasi-equilibrium at the current value of $\chi$. Note that, in writing Eq.(\ref{rho-chi}), we already are invoking the second law of effective-temperature thermodynamics.  

The first law of thermodynamics for this system is
\begin{equation}
\label{firstlaw}
\chi\,{\dot S_C\over V} = W - \bar W_S + Q.
\end{equation}
Here, the left-hand side is the rate of change of the configurational heat content per unit volume $V$.  On the right-hand side of Eq.(\ref{firstlaw}), $W = \dot\epsilon^{pl}\,\sigma$ is the rate at which inelastic external work is being done by the stress $\sigma$; and $Q$ is the (negative) rate at which heat is flowing from the kinetic-vibrational subsystem (the heat bath) into the configurational subsystem.  $\bar W_S$ is the rate at which the configurational internal energy $U_C$ is increasing, for example, by formation of new dislocations, at fixed configurational entropy $S_C$. This term is sometimes called the time derivative of the ``stored energy of cold work,'' but that expression can be misleading. Note that neither $\chi\,dS_C$ nor $\bar W_S\,dt$ are exact differentials; neither the configurational heat content nor the stored energy are state functions, and both incremental quantities include energies associated with dislocations and other internal, configurational degrees of freedom. 

Equation (\ref{firstlaw}) becomes an equation of motion for the effective temperature $\chi$ when we write it in the form
\begin{equation}
\label{firstlaw-chi}
c^{e\!f\!f}\,\dot\chi = W - \bar W_{\chi} +Q,
\end{equation}
where $c^{e\!f\!f}$ is the dimensionless, effective specific heat at constant volume and constant dislocation density $\rho$.  The quantity $\bar W_{\chi}$ is the rate at which $U_C$ is increasing at fixed $\chi$ instead of, as in $\bar W_S$, at fixed $S_C$.  The thermodynamic distinction between $\bar W_S$ and $\bar W_{\chi}$ was not made correctly in \cite{BLI-09} and \cite{BLII-09}.  It will not be important for present purposes; but there are other situations in which it becomes essential.  (We thank K. Kamrin for pointing out this earlier mistake to us.)  

To evaluate the terms in Eq.(\ref{firstlaw-chi}), we make two key observations.  First, because of the very large energies associated with creation and annihilation of dislocations, ordinary thermal fluctuations are completely ineffective in driving those processes.  Thus, the only scalar rate factor available to us is the rate $W$ at which work is being done by the driving force.  This quantity is non-negative because, in the absence of thermal fluctuations large enough to anneal out the dislocations or induce thermally assisted strain recovery, the strain rate must always be in the same direction as the stress. As a result, we may think of $W$ as being proportional to the non-negative strength of mechanically induced noise in an otherwise quiet environment.  It follows that each of the terms on the right-hand side of Eq.(\ref{firstlaw-chi}) is proportional to $W$.

Second, we know that, in steady-state deformation at dimensionless strain rate $q$, the effective temperature $\chi$ must approach a stationary value, say, $\chi_{ss}(q)$.  This function has been measured directly in molecular-dynamics simulations of glassy materials by Liu and coworkers. \cite{ONOetal-02,HAXTON-LIU-07} (See also \cite{EB-08}.) In the limit $q \ll 1$, where the shear rate is much smaller than any intrinsic rate in the system, they found that $\chi_{ss}(q\to 0)=\chi_0 $ is nonzero, roughly (perhaps exactly) equal to the glass transition temperature.  In other words, these systems reach fluctuating steady states of disorder under slow shear.  The slower the shear, the longer a system takes in real time to reach steady state; but the ultimate value of $\chi_{ss}$ in this limit is independent of $q$.  Liu et al. also observed that $\chi_{ss}(q)$ rises with increasing $q$ and appears to diverge as $q$ approaches unity.  We will need a model for that behavior when we study the limit of very high strain rates in Sec.\ref{VHSR}.  For the present, we simply assume that this function exists and is nonzero throughout the range $0 < q < 1$.     

Return now to the right-hand side of Eq.(\ref{firstlaw-chi}). The preceding argument about the rate factor implies that $\bar W_{\chi}$ is proportional to $W$.  We also know that the proportionality factor must be non-negative and less than or equal to unity, because only a fraction of the external work is converted to stored energy.  Finally, we know that $\bar W_{\chi}$ vanishes in steady state, where all of the work done is dissipated as heat, $W = -\, Q$, and where $\chi = \chi_{ss}$. Thus, for values of $\chi$ not too far from $\chi_{ss}$, we can make a linear approximation and write
\begin{equation}
\label{Ws}
\bar W_{\chi} = (1- \kappa)\,\left(1- {\chi\over \chi_{ss}}\right)\,W,~~~~0 < \kappa < 1,
\end{equation}
where $\kappa$ is a system-specific parameter that is proportional to the fraction of the external work that is converted into configurational heat. The parameter $\kappa$ will play a central role in the discussion of strain hardening in Sec.\ref{SH1}.

Using the same reasoning, we can write the heat flow in the form
\begin{equation}
\label{Q}
Q = -\, W\,{\chi\over \chi_{ss}(q)}.
\end{equation}
The factor $\chi$ appears here because a conventional heat-flow is proportional to the temperature difference $k_B\,T - \chi$, and we know that $k_B\,T \ll \chi$.  The factor $\chi_{ss}^{-1}$ ensures that $\chi_{ss}$ retains its meaning as the steady state value of $\chi$.  Note that Eq.(\ref{Q}) predicts that the differential heating coefficient, usually denoted by the symbol $\beta_{diff} \equiv - Q/W$, is simply equal to $\chi/\chi_{ss}$.  This formula looks qualitatively like the data shown in \cite{MACDOUGALL-00}; it merits further investigation.  

Equation (\ref{firstlaw-chi}) now becomes
\begin{equation}
\label{chidot}
c^{e\!f\!f}\,\dot\chi = \kappa\,\dot\epsilon^{pl}\,\sigma\,\left[1-{\chi\over \chi_{ss}(q)}\right].
\end{equation}
It is convenient to rewrite this equation using the total strain $\epsilon$ instead of the time as the independent variable:
\begin{equation}
\label{chi-epsilon-0}
{d\chi\over d\epsilon} = \kappa\,\sigma\,{q(\sigma,\rho)\over c^{e\!f\!f}\,q_0}\,\left[1-{\chi\over \chi_{ss}(q)}\right];~~~~ q_0 = \dot\epsilon\,\tau_0.
\end{equation} 
Equation (\ref{chidot}) is closely related to its STZ-theory analog, Eq.(5.10) in \cite{BLIII-09}. The detailed derivation presented in that paper includes the effects of ordinary thermal fluctuations, whose absence simplifies the present result. 

\section{Strain hardening and an equation of motion for the dislocation density}
\label{SH1}

We now need equations of motion for the stress $\sigma$ and the dislocation density $\rho$, relating both to the effective temperature $\chi$ as determined by Eqs.(\ref{chidot}) or (\ref{chi-epsilon-0}).  

To obtain an equation for the stress, assume that the total strain rate, $\dot\epsilon$, is the sum of a plastic part $\dot\epsilon^{pl}$ given by Eq.(\ref{qdef}), and an elastic part $\dot\sigma/\mu$, where $\mu$ is the shear modulus.   Then, converting to total strain $\epsilon$ as the independent variable, we have
\begin{equation}
\label{sigma-epsilon}
{d\sigma\over d\epsilon} = \mu\,\left[1 - {q(\sigma,\rho)\over q_0}\right];~~~~ q_0 = \dot\epsilon\,\tau_0,
\end{equation}
where $q(\sigma,\rho)$ is given by Eq.(\ref{qdef}) supplemented by Eq.(\ref{Taylor}) for the Taylor stress as a function of $\rho$.  

In writing an equation of motion for $\rho$, we run into a serious problem in the conventional literature. Much of modern dislocation theory is based on the so-called ``storage-recovery equation'' first proposed by Kocks and collaborators \cite{KOCKS-66,MECKING-KOCKS-81,FOLLANSBEE-KOCKS-88,KOCKS-MECKING-03}:
\begin{equation}
\label{rhoeqn}
d\rho/d\epsilon = k_1\,\sqrt{\rho} - k_2\,\rho,
\end{equation}
where $k_1$ and $k_2$ are $\rho$-independent parameters. Here, $\rho$ is meant to be the density of just the stored (i.e. pinned) dislocations; and $\sqrt{\rho}$ is the inverse of the dislocation spacing $\ell$, which is assumed to be the mean free path for mobile dislocations.  The first term on the right-hand side of Eq.(\ref{rhoeqn}) is said to be a storage rate, and the second term is the rate of annihilation or mobilization of stored dislocations.  Equation(\ref{rhoeqn}) is the starting point for a large part of the phenomenological literature in this field, including \cite{KUBIN-08} and \cite{PTW-03}.  If we say that the stress is always equal to the Taylor stress, and replace $\sqrt{\rho}$ by a term proportional to $\sigma$, we obtain the Voce equation \cite{VOCE-47}, which implies that $\sigma(\epsilon)$ rises linearly from zero and relaxes exponentially to a steady-state flow stress in the limit of large, positive $\epsilon$.  With various modifications and the addition of other parameters, stress-strain curves of this kind can be made to fit a wide range of experimental data. 

One problem with Eq.(\ref{rhoeqn}) is that the left-hand side changes sign when the shear rate changes direction; thus, this equation violates time reversal and reflection symmetries. The symmetry problem can be solved superficially by using only positive values of the strain, effectively by replacing $d\rho/d\epsilon$ by its absolute value; but such a mathematical singularity cannot appear at this place in any physically well posed equation of motion. Eq.(\ref{rhoeqn}), or the equivalent Voce equation as generalized by PTW, provides a phenomenological curve-fitting device;  but it can have no predictive value of its own.

In fact, important physics is missing.  This phenomenological approach provides no way to introduce and test any specific physical model such as the one proposed here in Sec.\ref{Model-dynamics}.  Moreover, although Eq.(\ref{rhoeqn}) supposedly describes the evolution of $\rho$, it contains no connection between the rate at which mechanical work is being done on the system and the rate at which dislocations are being created or annihilated.  Nothing happens in this system unless work is being done on it, so it seems obvious that the equation of motion for $\rho$ -- in analogy to the equation for $\chi$ -- must contain the external forcing. 

Accordingly, we propose to discard Eq.(\ref{rhoeqn}) entirely, and replace it by an equation based on the second law of thermodynamics and energy-conservation.  The second law requires that $\rho$ relax toward its most probable value $\rho^{ss}(\chi)$.  Therefore, for $\rho$ not too far from $\rho^{ss}$, we write the equation of motion for $\rho$ in the form
\begin{equation}
\label{rho-epsilon}
{d\rho\over d\epsilon}= {\kappa_{\rho}\over \gamma_D}\,{\sigma\,q(\sigma,\tilde\rho)\over q_0}\,\left[1 - {\rho\over \rho^{ss}(\chi)}\right];~~~~ q_0 = \dot\epsilon\,\tau_0.
\end{equation}
Here $\kappa_{\rho}$ is a dimensionless energy-conversion coefficient analogous to $\kappa$ in Eqs.(\ref{chidot}) and (\ref{chi-epsilon-0}), and $\gamma_D$ is the energy per unit length of a dislocation.  Thus, the prefactor in Eq.(\ref{rho-epsilon}) is the rate at which dislocations are being created if a fraction $\kappa_{\rho}$ of the work done on the system is stored in that form. Note that this prefactor is non-negative in accord with the second law.  The work rate $\sigma\,q$ is a non-negative scalar, and $q_0$ changes sign when $\dot\epsilon$ changes sign. 

Formally, Eq.(\ref{rho-epsilon}) can be interpreted as having emerged from a Clausius-Duhem inequality requiring that the rate of entropy production be non-negative.  It is shown in \cite{BLI-09,BLII-09} that such inequalities, and equations of motion of this form, follow directly from the principle that the statistically defined entropy must be a non-decreasing function of time.  (A similar equation was used in \cite{BLP07II} to describe the approach to quasi-equilibrium of a density-like variable.) By invoking this principle, we eliminate the need at this point in the analysis to specify the mechanisms by which the system achieves steady state, i.e. ``dynamic recovery.''  In Cottrell's terms \cite{COTTRELL-02}, we can ``interpret'' this process as being achieved by a balance between many dislocation creation, annihilation, and transformation mechanisms; but, for ``predictive'' purposes, we need only a few parameters that somehow contain all of this mechanistic information.  We already have seen that the activation energy $k_B\,T_P$ in Eq.(\ref{UP}) can describe a wide variety of rate-limiting processes.  In Eqs.(\ref{chidot}) and (\ref{rho-epsilon}), the energy-conversion coefficients $\kappa$ and $\kappa_{\rho}$ are playing similar roles.  

According to the discussion preceding Eq.(\ref{chidot}), $\kappa$ is proportional to the fraction of the work of deformation that is converted into configurational heat.  We expect $\kappa$ to be a  temperature dependent quantity.  At low $T$, the work of deformation goes primarily into producing new dislocations, which means that it is stored in the form of recoverable energy, and, according to Eq.(\ref{Ws}), $\kappa$ is small.  At higher temperatures, where pinning forces are weaker and the system can explore a larger range of configurations, more entropy is generated, and thus more of the energy is converted into configurational disorder.  Therefore, we expect $\kappa$ to increase with increasing temperature.  As yet, however, we have no way to compute  $\kappa$ from first principles; therefore, we use it as an adjustable parameter.   

The conversion coefficient $\kappa_{\rho}$ necessarily contains more structure, because it governs, not just the rate at which $\rho$ approaches its steady-state equilibrium value, but also the initial hardening rate. The onset of hardening is governed by Eq.(\ref{sigma-epsilon}), which is a stiff differential equation because the shear modulus $\mu$ is about two orders of magnitude larger than the Taylor stress.  The crossover from the initial elastic behavior, where $\sigma \cong \mu\,\epsilon$, to the onset of hardening occurs when $q$ becomes nearly equal to $q_0$. This happens at a value of the strain that is of the order of $10^{-5}$ in the examples to be described in Sec.\ref{SH2}.  Beyond that strain, $q$ remains essentially constant at $q_0$ -- the strain rate becomes  entirely plastic -- and the stress required to maintain the fixed $q = q_0$ grows because the density of dislocations is growing in accord with Eqs.(\ref{chi-epsilon-0}) and (\ref{rho-epsilon}).  Thus, hardening is a slow and very nearly steady-state process.  To a good approximation, we can use Eq.(\ref{sigma-sigmaT}), with $\nu(T,\,\rho,\,q)$ replaced by $\nu(T,\,\rho,\,q_0)$, to deduce that the ratio $\sigma/\sigma_T$ is equal to a weakly temperature and strain-rate dependent constant throughout the hardening process.  

Now consider the onset of hardening, where the total strain rate is changing from elastic to plastic, and the plastic deformation is just becoming visible on a stress-strain graph.  Experimental evidence (see, for example, Fig.21 in \cite{KOCKS-MECKING-03} or Fig.3 in \cite{FOLLANSBEE-KOCKS-88}) indicates that the initial hardening rate, in units of the shear modulus, is roughly independent of both temperature and strain rate, and has a magnitude
\begin{equation}
\label{Theta0}
{\Theta_0\over \mu} \equiv {1\over \mu}\,\left({d\sigma\over d \epsilon}\right)_{\rm onset} \cong {1\over 20}.
\end{equation} 
The question is how to incorporate this experimentally observed, apparently universal, onset condition into the equation of motion for $\rho$.  One possibility might be to supplement the activated depinning rate in Eqs.(\ref{vP}) and (\ref{fP}) by some temperature and strain-rate independent mechanism; but that procedure would move us away from our strategy of exploring only the minimal model introduced in Sec.\ref{Model-dynamics}.  For the present, therefore, we reserve the possibility of alternative dynamical mechanisms for later investigation, and simply assume here that the onset physics is contained in the conversion coefficient $\kappa_{\rho}$.  

To implement this assumption, look at Eq.(\ref{rho-epsilon}) in the case where the system is  beyond onset but $\rho$ is still much less than $\rho^{ss}$.  In this case, we know that $q = q_0$, and that $\sigma \cong \nu_0\,\sigma_T$, where $\nu_0 = \nu(T,\,\rho,\,q_0)$. Therefore
\begin{equation}
\label{onset1}
\left({d \rho\over d \epsilon}\right)_{\rm onset} \cong {\kappa_{\rho}\,\nu_0\,\sigma_T\over \gamma_D} = {b\,\,\kappa_{\rho}\,\nu_0\,\mu_T\over \gamma_D}\,\sqrt{\rho}.
\end{equation}
Solving for $\rho$, we find
\begin{equation}
{d\sigma\over d\epsilon} \cong {\nu_0^2\,\mu_T^2\,b^2\over 2\,\gamma_D}\,\kappa_{\rho}.
\end{equation}
If we set the right-hand side of this relation equal to $\Theta_0$ and use the approximation in Eq.(\ref{Theta0}), the prefactor on the right-hand side of Eq.(\ref{rho-epsilon}) becomes 
\begin{equation}
{\kappa_{\rho}\over \gamma_D} = {2\,\Theta_0\over (\nu_0\,\mu_T\,b)^2} \approx {\mu\over 10\,(\nu_0\,\mu_T\,b)^2},
\end{equation}
and we have completely determined all the parameters in Eq.(\ref{rho-epsilon}).

This procedure evades the question of why or whether $\Theta_0/\mu$ can be a universal ratio, independent of strain rate or temperature, for a wide range of experimental situations.  In this connection, note that -- {\it if} -- we assume that onset always occurs when $\sigma$ is some fixed multiple of $\sigma_T$, and that a fixed fraction of the work of deformation always is converted into new dislocations at this point, then energy conservation implies a strain-rate-independent relation of the form of Eq.(\ref{onset1}):
\begin{equation}
\label{onset2}
\left({d \rho\over d \epsilon}\right)_{\rm onset} \propto {\sigma_T\over \gamma_D} = {\mu_T\over \gamma_D}\,\sqrt{\rho}.
\end{equation}
Thus,
\begin{equation}
\label{Theta02}
{\Theta_0\over \mu} \approx {b^2\,\mu_T^2\over 2\,\mu\,\gamma_D},
\end{equation} 
which seems to be a plausible, temperature-independent estimate of $\Theta_0/\mu$ if $\mu$, $\gamma_D$, and $\mu_T$ all scale with temperature in the same ways.  More generally, however,  we expect that $\Theta_0$ is sensitive to many dynamical details and also to sample preparation, and therefore is not actually an intrinsic property of a material. 

\section{Parameters and Scaling}
\label{scaling}

Our next step is to identify conveniently rescaled variables and estimate values of parameters that emerge from this process. 

Because the formation energy $e_D$ in Eq.(\ref{rho-chi}) is large, and the effective temperature $\chi$ is the only energy in the theory that is comparable to it, we transform to the dimensionless ratio $\tilde\chi \equiv \chi/e_D$, and use this variable in the equations of motion.  There are several ways to estimate the scale of $\tilde\chi$ by purely geometric considerations.  Start with Eq.(\ref{rho-chi}) for the density of dislocations $\rho^{ss}$, and note that the length scale $a$ is the average spacing between dislocations when $\tilde\chi \to \infty$.  There is nothing unrealistic about the concept of an infinite $\tilde\chi$; it describes a state of maximum disorder in which any area $a^2$ is as likely to contain a dislocation as not to contain one.  (Infinite ``spin temperatures'' are commonly used to describe magnetic systems in which the moments are equally likely to be aligned parallel or antiparallel with some axis.)  Arbitrarily large values of $\tilde\chi$ play prominent roles in the high strain-rate analysis to be discussed in Sec.\ref{VHSR}.  The density $a^{-2}$, where $\tilde\chi \to \infty$, may be the value of $\rho$ where the interactions between dislocations are energetically comparable to their formation energies, so that the Boltzmann approximation breaks down.  Thus, it seems reasonable to guess that $a$ might be about ten atomic spacings.  

Now recall that $\tilde\chi_0 \equiv \tilde \chi_{ss}(q \to 0)$, via Eq.(\ref{rho-chi}), sets the density of dislocations when the system undergoes arbitrarily slow deformations for arbitrarily long times.  We propose that this definition of $\tilde\chi_0$ be interpreted very roughly as a system-independent geometric criterion, weakly analogous to the idea that amorphous materials become glassy when their densities are of the order of maximally random jammed packings, or to the Lindemann criterion according to which crystals melt when thermal vibration amplitudes are of the order of a tenth of the lattice spacing.  In that spirit, we guess that $\tilde\chi_0$ is the dimensionless effective temperature at which the spacing between dislocations $\ell$ is roughly ten length scales $a$, or about one hundred atomic spacings.  This estimate is consistent with the dislocation spacings at the upper end of the graph of $\sqrt{\rho}$ versus stress shown in Fig.1 of Mecking and Kocks \cite{MECKING-KOCKS-81}.  Thus we guess that $1/\tilde\chi_0 \sim 2\,\ln (10) \sim 4$ and, from here on, use $\tilde\chi_0 = 0.25$. In the future, when we consider more complex models containing multiple energy scales comparable to $e_D$, we may be able to obtain a better estimate of $\tilde\chi_0$.  For the moment, however, we presume that any inaccuracy in this estimate is compensated by variations in other parameters such as the ratio $b/a$. 

\begin{figure}[h]
\centering \includegraphics[height=7 cm]{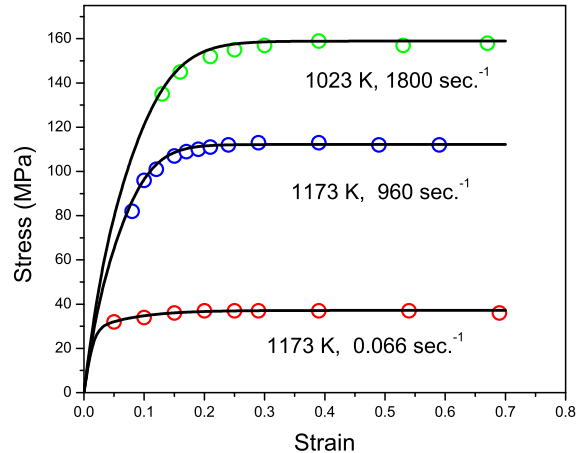}
\caption{\label{TD-1}  (Color online) Stress-strain graphs for Cu at relatively high temperatures as shown.  The data points are taken from PTW \cite{PTW-03}, Fig.2.  Temperatures and strain rates are shown on the graph.  The theoretical curves are computed with $T_P = 40,800\,K$ and $\tilde\chi_{ss} = \tilde\chi_0 = 0.25$. For $T = 1173\,K$, $\bar\mu_T = 1343$ MPa, and ${\cal K} = 0.11\,({\rm MPa})^{-1}$.  The initial values of $\tilde\chi$ are $\tilde\chi_{in} = 0.18$ for the upper curve with large strain rate, and $\tilde\chi_{in} = 0.22$ for the bottom curve, where the small strain rate apparently allows $\tilde\chi_{in}$ to be close to its steady-state value. For $T= 1023\,K$, $\bar\mu_T = 1490$ MPa, ${\cal K} = 0.055\,({\rm MPa})^{-1}$, and $\tilde\chi_{in}= 0.185$. }
\end{figure}

This rescaling of the effective temperature suggests that we define
\begin{equation}
\label{rho-chi-2}
\tilde{\rho}(\tilde\chi)\equiv a^2\,\rho(\chi); ~~~~\tilde\rho^{ss}(\tilde\chi) = e^{-\,1/\tilde\chi}.
\end{equation}
Then we rewrite Eq.(\ref{qdef}) in the form
\begin{equation}
\label{qdef2}
q(\sigma,\rho) \equiv \dot\epsilon^{pl}\,\tau_0 = (b/a)\,\tilde q(\sigma,\tilde\rho),
\end{equation}
where
\begin{equation}
\label{qtilde}
\tilde q(\sigma,\tilde\rho) =\sqrt{\tilde\rho}\,\Bigl[f_P(\sigma)-f_P(-\sigma)\Bigr].
\end{equation}
Equation (\ref{sigma-sigmaT}) becomes
\begin{equation}
\label{sigma-sigmaT-2}
{\sigma\over \sigma_T} \approx \ln\left({T_P\over T}\right) - \ln\left[{1\over 2}\,\ln\left({\tilde\rho\over \,\tilde q^2}\right) \right]\equiv \tilde\nu(T,\,\tilde\rho,\,\tilde q).
\end{equation}
The Taylor stress is 
\begin{equation}
\label{Taylor2}
\sigma_T = \bar\mu_T\,\sqrt{\tilde\rho},~~~~\bar\mu_T \equiv (b/a)\mu_T.
\end{equation}
We also use the definition of $\tilde q$ in Eq.(\ref{qtilde}) to rescale the steady-state effective temperature:
\begin{equation}
\label{tildeq-ss}
\tilde\chi_{ss}(\tilde q) \equiv \chi_{ss}(q)/e_D.
\end{equation}

Using $\tilde q$ instead of $q$ as the dimensionless measure of plastic strain rate means that we are effectively rescaling $\tau_0$ by a factor $b/a$.  For purposes of this analysis, we assume that $(a/b)\,\tau_0 = 10^{-12}\,{\rm sec.}$, independent of temperature; and we use $\tilde q = 10^{-12}\,\dot\epsilon^{pl}$ for converting from $\tilde q$ to measured strain rates.   This estimate of $\tau_0$ is about  the same as the atomic vibration time used by PTW.  The dimensionless rate $\tilde q $ is approximately equal to unity at the upper edge of the PTW data; and $\tilde q \ll 1$ throughout the thermal-activation region.  It follows that $\tilde\chi_{ss} = \tilde\chi_0 \cong 0.25$ in the latter region.  

\begin{figure}[h]
\centering \includegraphics[height=7 cm]{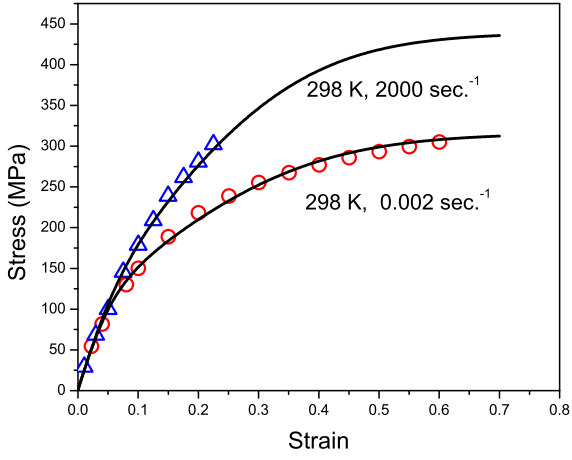}
\caption{\label{TD-2}  (Color online) Stress-strain graphs for Cu at $T = 293\,K$ and two different strain rates as shown. The data points are taken from \cite{FOLLANSBEE-KOCKS-88} and \cite{LANL-99}.  For both curves, $\bar\mu_T = 1600$ MPa,  ${\cal K} = 0.007\,({\rm MPa})^{-1}$, and $\tilde\chi_{in} = 0.18$. }
\end{figure}

We now use the steady-state flow stresses from the high-temperature stress strain curves for Cu shown in PTW Fig.2 -- shown here in Fig.\ref{TD-1} -- to compute $T_P$, which we assume to be independent of T.  We also use this information to evaluate the modulus $\bar\mu_T$ that appears in the definition of the Taylor stress in Eq.(\ref{Taylor2}).  In Eq.(\ref{sigma-sigmaT-2}), we set $\tilde\rho = \tilde\rho^{ss} = \exp\,(-1/\tilde\chi_0)$ with $\tilde\chi_0 = 0.25$.   Using just the steady-state flow stresses for the two curves at $T = 1173\,K$, with dimensionless rate factors $\tilde q = 9.6 \times 10^{-10}$ and $6.6 \times 10^{-14}$ we find $T_P = 40,800\,K$ and $\bar\mu_T = 1343$ MPa.  At the lower temperature, $T = 1023\,K$, we find $\bar\mu_T = 1490$ MPa.  At a yet lower temperature, $T = 298\,K$, using data shown in Fig.\ref{TD-2}, we find $\bar\mu_T = 1600$ MPa.

\section{Strain Hardening: Numerical Examples and Comparisons with Experiment}
\label{SH2}

In the scaled variables, the equations of motion that we need for the strain-hardening analysis are:
\begin{equation}
\label{sigma-epsilon-2}
{d\sigma\over d\epsilon} = \mu\,\left[1 - {\tilde q(\sigma,\tilde\rho)\over \tilde q_0}\right],~~~~ \tilde q_0 = (a/b)\,\dot\epsilon\,\tau_0;
\end{equation}
\begin{equation}
\label{chi-epsilon-2}
{d\tilde\chi\over d\epsilon} = {\cal K}\,\sigma\,{\tilde q(\sigma,\tilde\rho)\over \tilde q_0}\,\left[1 - {\tilde\chi\over\tilde\chi_{ss}(\tilde q)}\right],~~~~ {\cal K} \equiv {\kappa\over c^{e\!f\!f}\,e_D};
\end{equation}
and
\begin{equation}
\label{rho-epsilon-2}
{d\tilde\rho\over d\epsilon}= {{\cal K}_{\rho}\,\sigma\over \tilde\nu(T,\tilde\rho,\tilde q_0)^2}\,{\tilde q(\sigma,\tilde\rho)\over \tilde q_0}\,\left[1 - {\tilde\rho\over\tilde\rho^{ss}(\tilde\chi)}\right],
\end{equation}
where
\begin{equation}
\label{kapparho}
{\cal K}_{\rho} \equiv {2\,\Theta_0\over \bar\mu_T^2}\approx {\mu\over 10\,\bar\mu_T^2}.
\end{equation}
${\cal K}$ and ${\cal K}_{\rho}$ are temperature dependent quantities (the latter via $\mu$ and $\bar\mu_T$), with the dimensions of inverse stress.

\begin{figure}[h]
\centering \includegraphics[height=7 cm]{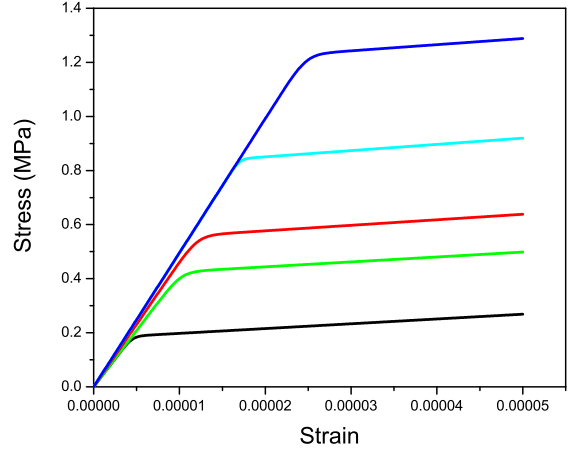}
\caption{\label{TD-Onset}  (Color online) Transition between elastic behavior and the onset of hardening at very small strains, for the five stress-strain curves shown in Figs.\ref{TD-1} and \ref{TD-2}. From bottom to top: $T=1173\,K$, $\dot\epsilon = 0.066\,{\rm sec}^{-1}$ (black);  $T=1173\,K$, $\dot\epsilon = 960\,{\rm sec}^{-1}$ (green); $T=1023\,K$, $\dot\epsilon = 1800\,{\rm sec}^{-1}$ (red); $T= 298\,K$, $\dot\epsilon = 0.002\,{\rm sec}^{-1}$(cyan); $T= 298\,K$, $\dot\epsilon = 2000\,{\rm sec}^{-1}$ (blue).}
\end{figure}

The experimental data sets on which we base our strain-hardening analyses are shown in Figs.\ref{TD-1} and \ref{TD-2}.  Both  figures show constant strain-rate, stress-strain curves for Cu, the first (taken from PTW \cite{PTW-03}, Fig.2) for two relatively high temperatures, the second (from \cite{FOLLANSBEE-KOCKS-88} and \cite{LANL-99}) at about room temperature. Note that, for both $T=1173\,K$ in Fig.\ref{TD-1} and $T=298\,K$ in Fig.\ref{TD-2}, the two strain rates shown differ by factors of about $10^6$.

The theoretical curves in both figures have been computed by integrating Eqs.(\ref{sigma-epsilon-2}), (\ref{chi-epsilon-2}), and (\ref{rho-epsilon-2}).  In Eq.(\ref{chi-epsilon-2}), we have set $\tilde\chi_{ss}(\tilde q) = \tilde\chi_0$ because we consider only $\tilde q \ll 1$. We have set $\Theta_0 = \mu/20$ as in Eq.(\ref{kapparho}), which is consistent with the data shown in Fig.\ref{TD-2}  and, in the absence of small-strain data in Fig.\ref{TD-1}, provides a plausible extrapolation to zero strain.  We have assumed that $\mu \cong 50$ GPa at $T=298\,K$, and that $\bar\mu_T$ scales like $\mu$ as a function of temperature.  Thus, at all temperatures, we have used the low-temperature ratio $\mu \cong 31\,\bar\mu_T$ and, in evaluating the right-hand side of Eq.(\ref{kapparho}), have written ${\cal K}_{\rho} \cong 3.1/\bar\mu_T$. 

 For initial conditions, at $\epsilon = 0$, we have used $\sigma(0) =  0$, and have arbitrarily chosen a non-zero initial value of the dislocation density, $\tilde\rho(0) = 10^{-7}$, because we are not especially interested in the earliest stages of deformation of ultra-pure crystals.  The initial value of $\tilde\rho$ determines the stress at which the onset of hardening occurs, but seems to have little or no effect on the subsequent hardening so long as $\tilde\rho(0) \ll \tilde\rho^{ss}$. Accordingly, the only  parameters that we have varied freely to fit the data (having determined values of $\bar\mu_T$ from steady-state flow stresses) are ${\cal K}$ and the initial value of $\tilde\chi$, say, $\tilde\chi_{in}$.  The values of these parameters are shown in the figure captions.  As expected, ${\cal K}$ is independent of strain rate, and decreases substantially with decreasing temperature.  

\begin{figure}[h]
\centering \includegraphics[height=7 cm]{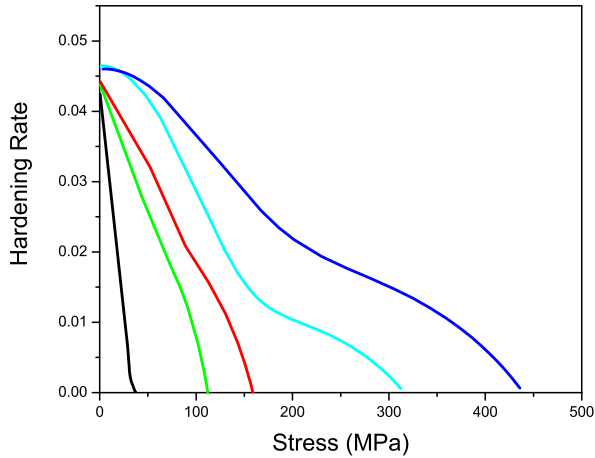}
\caption{\label{TD-3}  (Color online) Dimensionless hardening rate $\Theta(\sigma)/\mu$ for the five stress-strain curves shown in Figs.\ref{TD-1} and \ref{TD-2}. The color code is the same as in Fig.\ref{TD-Onset}, going left to right from higher to lower temperatures, and from smaller to larger strain rates.}
\end{figure}

The elastic regions near the origin are invisible in both figures. To show what is happening there theoretically, we zoom in to very small strains in Fig.\ref{TD-Onset}, where we see the expected elastic behavior, $\sigma = \mu\,\epsilon$, breaking off to the onset of hardening at stresses in the range $0.1 - 1.0$ MPa.  

The hardening rates,
\begin{equation}
{\Theta(\sigma)\over \mu} = {1\over \mu}\,{d\sigma\over d\epsilon}
\end{equation}
are shown in Fig.\ref{TD-3} for all five of the cases shown in Figs.\ref{TD-1} and \ref{TD-2}. At the lowest temperature, $T = 293\,K$ in Fig.\ref{TD-2} and the two right-most curves in Fig.\ref{TD-3}, the theoretical fits are almost insensitive to the initial effective temperature, $\tilde\chi_{in}$.  Here, hardening occurs in two stages roughly corresponding to stages II and III defined by Kocks and  Mecking in \cite{KOCKS-MECKING-03}. For a while after onset, the dislocation density grows in a deterministic way, independent of the effective temperature $\tilde\chi$, and is governed only by Eq.(\ref{rho-epsilon-2}) for $\tilde\rho \ll \tilde\rho^{ss}(\tilde\chi)$.  As is obvious both in the data and the theory, the early-stage hardening quickly becomes sensitive to strain rate and temperature; only the initial rate $\Theta_0$ is a strain-rate independent constant.  A later stage of hardening, starting near the inflection points on the right-most curves in Fig.\ref{TD-3}, sets in when $\tilde\rho$ becomes comparable in magnitude to $\tilde\rho^{ss}(\tilde\chi)$.  Then, hardening begins to be controlled by changes in $\tilde\chi$ as determined by Eq.(\ref{chi-epsilon-2}).  By this point in the process, however, $\tilde\chi_{in}$ is largely irrelevant, and only the rate at which $\tilde\chi$ approaches its steady-state value $\tilde\chi_0$ is important.

\begin{figure}[h]
\centering \includegraphics[height=7 cm]{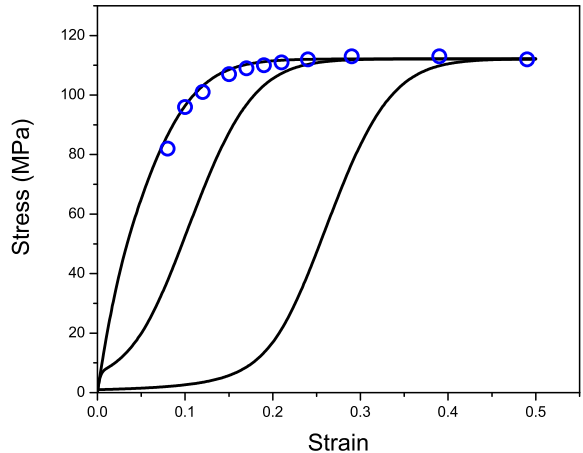}
\caption{\label{TD-chi-var}  (Color online) Effect of decreasing the initial value of $\tilde\chi$. All theoretical parameters are the same as those for the curve in Fig.\ref{TD-1} for $T=1173\,K$, $\dot\epsilon = 960\,{\rm sec.}^{-1}$, except that the values of $\tilde\chi_{in}$ are, from left to right, $0.18$, $0.10$, and $0.07$ respectively.  For reference, the original experimental data is indicated by the blue circles.} 
\end{figure}

The situation is quite different for the high-temperature behavior shown in Fig.\ref{TD-1} and by the three left-hand hardening curves in Fig.\ref{TD-3}.  Here, $\tilde\rho$ grows rapidly toward  $\tilde\rho^{ss}(\tilde\chi)$, and the dynamics of $\tilde\chi$ is already important at the earliest data points shown.  Since we have no data closer to onset, we are free to vary both ${\cal K}$ and $\tilde\chi_{in}$, subject only to the constraints that the theoretical curves join smoothly to the later-stage data, and that ${\cal K}$ should be independent of strain rate.  Our best-fit parameters seem to make the theory look as uninteresting as possible, but a wider range of behaviors is consistent with the data.  In fact, $\Theta(\sigma)$ is not a monotonically decreasing function if $\tilde\chi_{in}$ is small enough but is compensated by a larger value of ${\cal K}$. 

The wide range of hardening behaviors already emerging in this minimal model tells us that there is a rich variety of phenomena that can be predicted by analyses of this kind.  It must be possible to use the methods presented here to write the equations of motion for any well posed dynamical model of dislocations in a thermodynamically consistent way. As mentioned earlier, this procedure will involve introducing separate density functions for different kinds of dislocations.  

However, even in our minimal model, there are many phenomena that can occur.  For example, we show in Fig.\ref{TD-chi-var} what happens to the stress-strain curve for $T=1173\,K$, $\dot\epsilon = 960\,{\rm sec.}^{-1}$, if we fix ${\cal K}$ but decrease $\tilde\chi_{in}$ from its original value of $0.18$ to $0.10$ and then to $0.07$.  The resulting curves look qualitatively like those shown in Fig.1 of \cite{KOCKS-MECKING-03} for Cu single crystals oriented at different angles relative to the shear stress.  The similarity could be a coincidence; but it also could mean that the relevant populations of dislocations at each orientation have different formation energies and play different roles in the flow of energy and entropy through the system.  The thermodynamic analysis in Sec.\ref{Teff}, especially the relation between heat generation and the effective temperature in Eq.(\ref{Q}), might help to sort out these possibilities.

\section{Very high strain rates}
\label{VHSR}

We turn finally to the regime of high strain rates, specifically to the points derived from strong-shock data shown in Figs. 7 and 9 in PTW for steady-state values of $\dot\epsilon$ in the range $10^9$ to $10^{12}\,{\rm sec}^{-1}$.  With $\tau_0 \sim 10^{-12}\,{\rm sec}^{-1}$, our dimensionless strain rates $\tilde q$ are no longer much smaller than unity, and therefore we must pay attention to the $\tilde q$-dependence of $\tilde\chi_{ss}(\tilde q)$.  This issue has emerged prominently in studies of glassy systems, and has been investigated in detail in \cite{JSL-MANNING-TEFF-07}.  Much of the following discussion is based on the latter paper.

To estimate $\tilde\chi_{ss}(\tilde q)$ for large values of $\tilde q$, note that the inverse function, say $\tilde q = {\cal Q}(\tilde\chi_{ss})$, is a rate expressed as a function of a temperature.  As such, it can be compared to more familiar properties of disordered systems such as temperature dependent viscosities or diffusion constants.  For example, Liu et al \cite{ONOetal-02,HAXTON-LIU-07} found that their analog of $\tilde\chi_{ss}(\tilde q)$ increases rapidly with increasing $\tilde q$, and shows signs of diverging near $\tilde q \cong 1$. An Arrhenius plot of the low-temperature data in this range, shown in Fig. 1 of \cite{JSL-MANNING-TEFF-07}, reveals that 
\begin{equation}
\label{Q-chi}
{\cal Q}(\tilde\chi_{ss}) \approx e^{-\tilde e_A/\tilde\chi_{ss}},
\end{equation}
where $\tilde e_A = e_A/e_D$, and $e_A$ is an Arrhenius activation energy. For the glass simulations, $e_A$ is somewhat larger than the activation energy deduced directly from viscosity measurements (the analog of $e_D$).  The Arrhenius fit to Eq.(\ref{Q-chi}) in \cite{JSL-MANNING-TEFF-07} is much cleaner than is the corresponding fit to the simulated viscosity as a function of the bath temperature. 

It is tempting to speculate that there exists a universal, Arrhenius-like, steady-state relation between the effective temperature and the strain rate, and that this relation is valid for a wide class of athermal disordered systems including both glass formers and crystals with high densities of defects.  This speculation, of course, requires better theoretical justification and experimental tests before it can be accepted.  Note, however, that it is supported qualitatively by the high-strain-rate, steady-state data of PTW.  If we assume that the steady-state stress is always proportional to the Taylor stress, and use Eq.(\ref{Q-chi}) to evaluate $\tilde\rho^{ss} \sim \exp\,(-1/\tilde\chi_{ss})$ in Eq.(\ref{Taylor2}), we find
\begin{equation}
\label{q-beta}
\sigma \propto \bar\mu_T\,\tilde q^{\beta};~~~~ \beta = {e_D\over 2\,e_A}.
\end{equation}
The PTW results imply that $e_D/e_A \sim 0.5$, which is somewhat smaller than the value of this ratio ($\sim 0.67$) found for glass simulations in \cite{JSL-MANNING-TEFF-07}, but is not qualitatively inconsistent with it. In short, the effective temperature theory predicts that rate hardening is rapidly accelerated by growth in the density of dislocations when the strain rate becomes comparable to the intrinsic inverse time scale $\tau_0^{-1}$. 

To complete the theoretical development, we need an expression for $\tilde\chi_{ss}(\tilde q)$ that crosses over from $\tilde\chi_{ss}=\tilde\chi_0$ in the small-$\tilde q$ limit to the activated behavior shown in Eq.(\ref{Q-chi}).  Note that the easy-to-understand saturation of $\tilde\chi_{ss}$ at $\tilde\chi_0$ in the small-$\tilde q$ limit corresponds to the familiar -- yet mysterious -- divergence of the time scale at the glass transition.  In \cite{JSL-MANNING-TEFF-07} and elsewhere, an approximation for the rate factor in the whole range, $\tilde\chi_0 < \tilde\chi_{ss}(\tilde q) < \infty$, is written in the form
\begin{equation}
\label{chi-q-alpha}
{\cal Q}(\tilde\chi_{ss}) \approx \exp\left[-{\tilde e_A \over \tilde\chi_{ss}}- \alpha(\tilde\chi_{ss})\right],
\end{equation}
where $\alpha(\tilde\chi_{ss})$ is a Vogel-Fulcher function that has been modified so that it cuts off smoothly near $\tilde\chi_{ss} = \tilde\chi_A$, i.e. at the high-temperature end of the super-Arrhenius region:
\begin{equation}
\label{alpha}
\alpha(\tilde\chi_{ss}) = {\tilde\chi_1\over \tilde\chi_{ss} - \tilde\chi_0}\,\exp\left[- 3\,\left({\tilde\chi_{ss} - \tilde\chi_0\over \tilde\chi_A - \tilde\chi_0}\right)\right].
\end{equation}
(The factor $3$ is an arbitrary fitting parameter.)  This form of the rate factor emerges from the excitation-chain theory of the glass transition \cite{Langer-PRE-06,Langer-PRL-06}. A more physically motivated expression for $\alpha(\tilde\chi_{ss})$ has been introduced in \cite{JSL-AD-08}.

\begin{figure}[h]
\centering \includegraphics[height=7 cm]{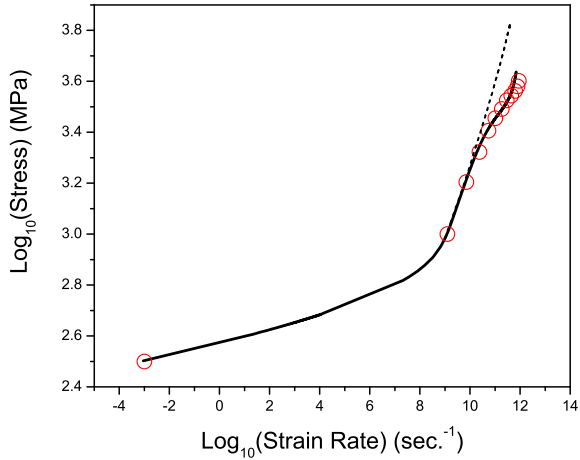}
\caption{\label{PTW-3}  (Color online) Stress versus strain rate for Cu at $T = 300\,K$.  The data points are taken from PTW \cite{PTW-03}, Fig.7. The solid theoretical curve is computed using $e_D/2\,e_A = 0.25$, $\tilde\chi_1=0.1$, $\tilde\chi_A=0.3$, and a thermal conductivity $K = 0.4$.  The dashed curve is computed without the thermal effect, i.e. with infinite $K$.}
\end{figure}

The dashed line in Fig.\ref{PTW-3} shows what happens when we set ${\cal Q}(\tilde\chi_{ss}) = \tilde q$, solve Eq.(\ref{chi-q-alpha}) for $\tilde\chi_{ss}(\tilde q)$, and insert the result into Eq.(\ref{sigma-sigmaT-2}) with $\tilde\rho = \tilde\rho^{ss} = \exp\,(-1/\tilde\chi_{ss})$ to compute the steady-state $\sigma$ as a function of $\tilde q$. The experimental points shown are taken directly from PTW Fig.7, for which the nominal temperature is $T = 300\,K$.  We also include the single point at $T=298\,K$, $\dot\epsilon = 2 \times 10^{-3}\,{\rm sec}^{-1}$ (the lower curve in Fig.\ref{TD-2}) in order to show the theoretical connection across the entire range of strain rates.  The best fit occurs at the PTW value, $e_D/2\,e_A = 0.25$. However, the actual slope of the dashed line in Fig.\ref{PTW-3} is closer to $0.3$, meaning that the approximations leading to the simple estimate in Eq.(\ref{q-beta}) -- especially dropping the $\tilde q$ and $\tilde\rho=\tilde\rho^{ss}$ dependences on the right-hand side of Eq.(\ref{sigma-sigmaT-2}) -- were not completely accurate. 

Note that $e_D/e_A$ is our only adjustable parameter; the quantities that determine $\alpha(\tilde\chi_{ss})$ in Eq.(\ref{alpha}) are irrelevant in this large-$q$ region, where ${\cal Q}(\tilde\chi_{ss})$ is dominated by the Arrhenius term.  Thus, our choice of $e_D/e_A$ determines, not only the slope of the curve at high strain rates, but also the location and shape of the crossover between the two different rate-hardening regimes.  There is very little flexibility in this fit.  The PTW estimate for $\beta$ was obtained, as shown in their Fig.9, by drawing a straight line through data whose slope is not constant. In fact, our predicted slope of approximately $0.3$ is correct for the first part of the high strain rate regime, but the experiments indicate that the rate-hardening effect weakens as the strain rate increases.

One possible explanation for this weakening is a thermal effect.  The activation rate introduced in Eq.(\ref{fP}) is a strongly temperature dependent quantity; small increases in $T$ increase this rate substantially.  According to Eq.(\ref{Q}), heat flows from the deforming configurational subsystem to the kinetic-vibrational subsystem at a rate $-\, Q$ equal, in steady state, to the rate at which work is done in driving the deformation.  If we suppose that the thermal conductivity between the configurational subsystem and an external heat bath is less than infinite, then the temperature $T$ must increase by an amount proportional to $Q$.  Specifically, we estimate that the heat flow is
\begin{equation}
\label{T-K}
Q = \tilde q\,\sigma = K\,(T - T_0),
\end{equation}
where $T_0$ is the ambient temperature and $K$ is a thermal conductivity. 

\begin{figure}[h]
\centering \includegraphics[height=7 cm]{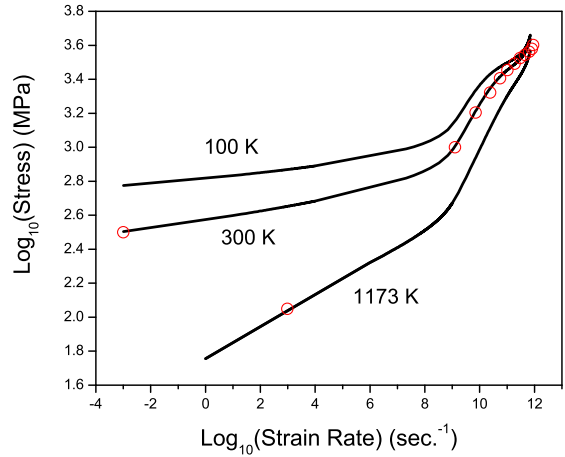}
\caption{\label{PTW-4}  (Color online) Stress versus strain rate for three different temperatures as shown on the graph. The theoretical parameters are the same as those used for Fig.\ref{PTW-3}.} 
\end{figure}
 
The solid curve in Fig.\ref{PTW-3} has been computed by inserting Eq.(\ref{T-K}) into Eq.(\ref{sigma-sigmaT}) and solving for $\sigma$ as a function of $\tilde q$  with $T_0 = 300\,K$,  $\beta = e_D/2\,e_A = 0.25$, and $K = 0.4$.  The agreement is remarkably good, considering the substantial uncertainties in both the experiments and the theory.  Among the theoretical uncertainties is the fact that Eq.(\ref{T-K}) predicts the temperature $T$ to be substantially larger than the melting point at the highest strain rates.  Perhaps the system can remain in a superheated solid state during the microscopically short time intervals over which the deformations take place.  (The experiments use shock fronts to achieve such high strain rates. These are not really steady-state measurements.)  Note also that the thermally modified theory still underestimates the thermal softening effect at the very highest strain rates, perhaps indicating that the material is indeed melting in that regime.   

To complete the comparisons with PTW, we show our version of their Fig.11 in Fig.\ref{PTW-4}.  Here, we supplement the $300\,K$ curve shown in Fig.\ref{PTW-3} with the stress-strain-rate curve for $1173\,K$ and include the experimental point deduced from the flow stress at $960\,{\rm sec}^{-1}$ as seen in Fig.\ref{TD-1}.  We also draw a curve for $T = 100\,K$ (guessing that $\bar\mu_T \cong 1700$ MPa), to illustrate that strain-rate hardening is extremely slow in general, and becomes slower at lower temperatures.  Even with the thermal effect included, there remains a substantial amount of temperature dependence at the highest strain rates.

\section {Summary and concluding remarks}
\label{summary}

The dislocations in a deforming crystalline or polycrystalline solid constitute a complex subsystem of the material as a whole.  The configurations of this subsystem have a wide range of energies, and there are extensive numbers of such configurations in any energy interval.  Such a system has an entropy, and therefore it has a temperature.  This ``effective'' temperature is enormously larger than the ordinary temperature of the solid. Although ordinary thermal fluctuations may affect the mobility of dislocations, only the work done by external forces is on a scale comparable to dislocation energies. Thus, to a first approximation, the subsystem of dislocations is decoupled from the heat bath, and the effective temperature is a well defined property of the subsystem.  The principal theme of this paper and its predecessors \cite{BLI-09,BLII-09,BLIII-09} is that any such driven subsystem is amenable to thermodynamic analysis. In particular, its equations of motion must be consistent with the first and second laws of thermodynamics.  

We have based our thermodynamic equations of motion primarily on two physical assumptions.  The first of these is the role of the Taylor stress in determining the rate of thermally assisted plastic flow, i.e. in Eqs.(\ref{Taylor}), (\ref{vP}), and (\ref{fP}).  The second assumption consists of a set of relations between the effective temperature and steady-state properties of the system, in particular, the quasi-equilibrium density of dislocations in Eq.(\ref{rho-chi}), and the plastic strain rate in Eq.(\ref{chi-q-alpha}). The former relation is simply a consequence of the fact that the effective temperature is the ``true'' temperature of the configurational subsystem.  The latter has emerged recently in theories of glass dynamics; its form may be generally valid for all strongly disordered, solidlike materials.  

Our theory contains only a small number of adjustable parameters, all of which have physical interpretations in terms of known properties of materials, and most of which are directly measurable by steady-state experiments. These parameters are: an overall microscopic time scale $\tau_0$; a baseline value of the dimensionless effective temperature $\tilde\chi_0$ that we estimate {\it a priori} from geometrical considerations; the height of the depinning barrier $T_P$; the shear modulus $\mu$; a reduced shear modulus $\bar\mu_T$ that determines the Taylor stress; the initial hardening rate $\Theta_0/\mu$ that we assume to be temperature and strain-rate independent; a ratio of characteristic dislocation energies $e_D/e_A$; and a quantity ${\cal K}$ that is proportional to the fraction of the work of deformation that is converted into configurational heat.  An analogous conversion factor, ${\cal K}_{\rho}$, related to formation of dislocations, is determined unambiguously by $\Theta_0$ and other known parameters.  Of these parameters, only ${\cal K}$ is a temperature (but not strain rate) dependent quantity that we choose to fit the transient, strain-hardening curves.  Our only other, similarly free, parameter is the initial value of the effective temperature, which necessarily depends on sample preparation.  With these ingredients, we predict strain hardening in agreement with experiment over a wide range of temperatures and strain rates. We also, with only the single adjustable parameter $e_D/e_A$, accurately predict both the crossover from moderate to very high strain rates and the power-law exponent $\beta$ observed in the latter regime.  

This theory provides a framework for further investigation in at least two important directions.  First, as stated at the end of Sec.\ref{SH2}, we should recast it in terms of a set of densities for different kinds of dislocations, each density function appearing as an internal state variable obeying its own dynamical equation of motion. (See \cite{BLI-09} for a discussion of the role of internal variables in nonequilibrium thermodynamics.)  The equations of motion for the different kinds of dislocations can include the mechanisms by which these populations interact with each other, thus possibly modifying the simple depinning model used throughout this paper.  In this extended development, we might also introduce other internal variables such as average grain size or impurity concentrations in order to make contact with other work in this field. We reiterate that it must be possible to use the methods presented here to write the equations of motion for any well-posed dynamical model of dislocations in a thermodynamically consistent way.  

Second, we should introduce spatial heterogeneities.  From the beginning of this analysis, we have considered only homogeneous systems; thus, we have ignored the formation of dislocation microstructures, shear-banding instabilities, fracture, and the like.  The reason for a lack of steady-state data at lower temperatures and higher strain rates is that materials fail heterogeneously under those conditions.  One  goal of a theory such as that proposed here is to predict quantitative failure criteria.  

In fact, the dynamical equations that we have derived are already continuum approximations; they can easily be generalized to situations in which $\sigma$, $\chi$ and $\rho$ are spatially varying fields.  For amorphous materials, the analogous effective-temperature theory already has provided a way to understand shear banding as a rate-weakening instability.  The effective heat generated in a region that happens to be deforming faster than its neighbors softens that region, accelerating the deformation rate and generating yet more heat.  Unlike the ordinary temperature, which diffuses very rapidly, the effective temperature hardly diffuses at all.  As a result, the effective-temperature theory predicts instabilities with realistic length and time scales.  See \cite{MANNINGetal-SHEARBANDS-08} for a comparison of this theory with molecular dynamics simulations of amorphous shear banding \cite{SHI-FALK-SHEARBANDS-07}, and \cite{MDLC-SHEARBANDS-09,DAUB-08} for applications to the dynamics of shear fracture in earthquake faults. 

An even wider range of dynamic heterogeneities is known to occur during dislocation-mediated deformation. It should be a fairly straightforward exercise to repeat the amorphous shear-banding analyses for polycrystalline solids using the present dislocation theory. The results clearly will be different because, unlike amorphous plasticity, simple dislocation mediated plasticity is rate strengthening. It will be interesting to learn what physical ingredients must be added to the present theory to produce strain localization. An equally interesting question is whether some version of this  theory can predict the diverse kinds of dynamic dislocation patterns discussed, for example, by Ananthakrishna. \cite{ANANTHAKRISHNA-07}


\begin{thebibliography}{99}

\bibitem{COTTRELL-53} A.H. Cottrell, {\it Dislocations and Plastic Flow in Crystals}, (Oxford University Press, London, 1953).

\bibitem{FRIEDEL-67} J. Friedel, {\it Dislocations} (Pergamon, Oxford, 1967).

\bibitem{HIRTH-LOTHE-68} J. Hirth and J. Lothe, {\it Theory of Dislocations}, (McGraw Hill, New York, 1968).

\bibitem{COTTRELL-02} A.H. Cottrell, in {\it Dislocations in Solids}, vol. 11, F.R.N. Nabarro, M.S. Duesbery, Eds. (Elsevier, Amsterdam, 2002), p. vii.  

\bibitem{KOCKS-MECKING-03} U.F. Kocks and H. Mecking, Prog. Matls. Sci. {\bf 48}, 171 (2003).

\bibitem{KUBIN-08} B. Devincre, T. Hoc and L. Kubin, Science {\bf 320}, 1745 (2008).

\bibitem{COLEMAN-NOLL-63} B. D. Coleman and W. Noll, Archive for Rational Mechanics and Analysis {\bf 13}, 167 (1963).

\bibitem{COLEMAN-GURTIN-67} B. D. Coleman and M. E. Gurtin, J. Chem. Phys. {\bf 47}, 597 (1967).

\bibitem{BLI-09} E. Bouchbinder and J.S. Langer, Phys. Rev. E {\bf 80}, 031131 (2009).

\bibitem{BLII-09} E. Bouchbinder and J.S. Langer, Phys. Rev. E {\bf 80}, 031132 (2009).

\bibitem{BLIII-09} E. Bouchbinder and J.S. Langer, Phys. Rev. E,{\bf 80}, 031133 (2009).

\bibitem{ZAISER-06} M. Zaiser, Advances in Physics {\bf 55}, 185 (2006).

\bibitem{FL-98}  M. L. Falk and J. S. Langer, Phys. Rev. E {\bf 57}, 7192 (1998).

\bibitem{JSL-STZ-PRE-08} J. S. Langer, Phys. Rev. E {\bf 77}, 021502 (2008).

\bibitem{PTW-03} D.L. Preston, D. L. Tonks, and D.C. Wallace, J. Appl. Phys. {\bf 93}, 211 (2003).

\bibitem{KOCKS-66} U.F. Kocks, Phil. Mag. {\bf 13}, 541 (1966).

\bibitem{MECKING-KOCKS-81} H. Mecking and U.F. Kocks, Acta Metall. {\bf 29}, 1865 (1981).

\bibitem{FOLLANSBEE-KOCKS-88} P.S. Follansbee and U.F. Kocks, Acta Metall. {\bf 36}, 81 (1988).

\bibitem{ANANTHAKRISHNA-07} G. Ananthakrishna, Physics Reports {\bf 440}, 113 (2007).

\bibitem{JSL-MANNING-TEFF-07} J. S. Langer and M. L. Manning, Phys. Rev. E {\bf 76}, 056107 (2007).

\bibitem{STILLINGER-WEBER-82} F. H. Stillinger and T. A. Weber, Phys. Rev. A {\bf 25}, 978 (1982).

\bibitem{STILLINGER-88} F. H. Stillinger, J. Chem. Phys. {\bf 88}, 7818 (1988).

\bibitem{ONOetal-02} I. K. Ono, C. S. O'Hern, D. J. Durian, S. A. Langer, A. Liu and S. R. Nagel, Phys. Rev. Lett. {\bf 89}, 095703 (2002).

\bibitem{HAXTON-LIU-07} T. Haxton and A. J. Liu, Phys. Rev. Lett. {\bf 99}, 195701 (2007).

\bibitem{EB-08} E. Bouchbinder, Phys. Rev. E {\bf 77}, 021502 (2008).

\bibitem{MACDOUGALL-00} D. Macdougall, Experimental Mechanics {\bf 40}, 306, 2000.

\bibitem{VOCE-47} E. Voce, J. Inst. Met. {\bf 74},537 (1947/48).

\bibitem{BLP07II} E. Bouchbinder, J. S. Langer and I. Procaccia, Phys. Rev. E {\bf 75}, 036108 (2007).

\bibitem{LANL-99} S.R. Chen, P.J. Maudlin, and G.T. Gray, III, ``Constitutive Behavior of Model FCC, BCC, and HCP Metals: Experiments, Modeling and Validation,'' pp. 623-626, in {\it The Seventh International Symposium on Plasticity and Its Current Applications}, A.S. Khan, ed. (Cancun, Mexico, Neat Press, 1999).

\bibitem{Langer-PRE-06} J. S. Langer, Phys. Rev. E {\bf 73}, 041504 (2006)

\bibitem{Langer-PRL-06} J. S. Langer, Phys. Rev. Lett. {\bf 97}, 115704 (2006).

\bibitem{JSL-AD-08} J.S. Langer, Phys. Rev. E {\bf 78}, 051115 (2008)

\bibitem{MANNINGetal-SHEARBANDS-08} M. L. Manning, J. S. Langer and J. M. Carlson, Phys. Rev. E {\bf 76}, 056106 (2007).

\bibitem{SHI-FALK-SHEARBANDS-07} Y. Shi, M. B. Katz, H. Li, and M. L. Falk, Phys. Rev. Lett. {\bf 98},
185505 (2007).

\bibitem{MDLC-SHEARBANDS-09} M.L. Manning, E.G. Daub, J.S. Langer, and J.M. Carlson, Phys. Rev. E {\bf 79}, 016110 (2009).

\bibitem{DAUB-08} E.G. Daub, M. L. Manning and J. M. Carlson, Geophys. Res. Letts. {\bf 35}, L12310, (2008).



\end{thebibliography}
\end{document}